\documentclass[twocolumn,aps,showpacs,superscriptaddress,prstab]{revtex4}
\pdfoutput=1 
\usepackage{amsmath}
\usepackage{amsfonts}
\usepackage{amssymb}
\usepackage{graphicx}
\usepackage{graphicx}
\usepackage{dcolumn}
\usepackage{bm}
\usepackage[sort&compress]{natbib}

\bibpunct{[}{]}{,}{n}{,}{,}

\begin{document}

\title{Beam quality requirements for the Ion-Channel Laser}

\author{X. Davoine}
\email{xavier.davoine@cea.fr}
\affiliation{
GoLP/Instituto de Plasmas e Fus\~ao Nuclear, Instituto Superior T\'ecnico, Lisbon, Portugal
}
\affiliation{
CEA DAM DIF, 91297 Arpajon, France
}

\author{F. Fi\'uza}
\affiliation{
GoLP/Instituto de Plasmas e Fus\~ao Nuclear, Instituto Superior T\'ecnico, Lisbon, Portugal
}

\author{R. A. Fonseca}
\affiliation{
DCTI/ISCTE -- Lisbon University Institute, Lisbon, Portugal
}
\affiliation{
GoLP/Instituto de Plasmas e Fus\~ao Nuclear, Instituto Superior T\'ecnico, Lisbon, Portugal
}

\author{W. B. Mori}
\affiliation{
University of California Los Angeles, Los Angeles, USA
}

\author{L. O. Silva}
\affiliation{
GoLP/Instituto de Plasmas e Fus\~ao Nuclear, Instituto Superior T\'ecnico, Lisbon, Portugal
}

\date{\today}

\begin{abstract} 
In this paper, we determine the electron beam quality requirements to obtain exponential radiation amplification in the ion-channel laser, where a relativistic electron beam wiggles in a focusing ion-channel that can be created in a wakefield accelerator.
The beam energy and wiggler parameter spreads should be limited.
Those spread limits are functions of the Pierce parameter, which is calculated here without neglecting the radiation diffraction.
Two dimensional and three dimensional simulations of the self-consistent ion-channel laser confirm our theoretical predictions.
\end{abstract}

\pacs{41.60.Cr,52.59.Ye,52.65.Rr,52.38.Kd}

\maketitle

The Ion-Channel Laser (ICL) \cite{whit90} relies on the injection of a relativistic electron beam in an Ion-Channel (IC) to create a coherent and highly amplified radiation source.
Such an IC can be produced in a plasma-based wakefield accelerator in the blowout or bubble regime \cite{mang04,gedd04,faur04}: while propagating in a plasma, a laser pulse or a particle beam pushes the electrons off-axis and lets an IC in its wake.
The fields generated in the IC provide a focusing force for the relativistic electrons on-axis. 
The resulting wiggling motion of the electron along the IC axis then leads to the emission of the so-called betatron radiation \cite{esar02,rous04}.
For appropriate conditions, betatron radiation can interact with the electron beam and bunch it at the radiation wavelength, allowing for the exponential amplification of the emitted radiation, like in a conventional FEL.
One of the most important advantages of the ICL are the strong fields generated in the plasma, which can lead to amplification in the UV to X-ray range with very high brightness within much shorter distances than those obtained in the conventional FEL sources.
Previous works analyzed the ICL gain length and the associated Pierce parameter assessments \cite{whit90,chen90,whit92,cliu07}.

In order to take full advantage of this scheme, it is critical to understand what are the requirements in terms of the beam quality to obtain high-gain, since the focusing structure is easily determined solely by the plasma density and the radius of the blowout/bubble region.
In this paper, we present a detailed analysis of the beam requirements when the Pierce parameter $\rho$ is much smaller than 1, as required for FEL-like amplification.
In an ICL, the wiggler parameter $K$ depends on the electron properties, so it can be different for each electron. Therefore, we show that both the beam energy spread and beam wiggler parameter spread should be limited and satisfy:
\begin{equation}
\frac{\Delta\gamma}{\gamma}<\frac{2}{3}\rho\ \ \ {\rm and}\ \ \ \frac{\Delta K}{K}<\frac{2+K^2}{2K^2}\rho
\label{ICL_working_conditions}
\end{equation}
Multi-dimensional PIC simulations of ICL are performed to confirm that if those conditions are fulfilled then a good amplification is observed.
As the spread limitations are a function of the Pierce parameter, this parameter should be carefully calculated.
However, two important effects were neglected in previous works \cite{whit90,chen90,whit92,cliu07}: (i) the radiation diffraction and (ii) the Pierce parameter dependence on the wiggler parameter $K$.
These effects are included in our theoretical calculation of the Pierce parameter and the associated gain length, and are confirmed by PIC simulations in Lorentz boosted frames.

As a first step, we analyze the motion of an electron in an IC whose boundary is described by a radius, $r_b$ which depends on the variable, $\xi=z-ct$. 
In general the motion of particle moving near the speed of light in an IC can be described in terms to the so-called wake potential $\psi \equiv \frac{e}{mc^2} (\phi-A_z)$ where $\phi$ and $A_z$ are the scalar potential and axial component of the vector potential. 
The accelerating and focusing fields are obtained from $\frac{\partial}{\partial\xi} \psi$ and $\frac {\partial}{\partial r} \psi$ where we assume azimuthal symmetry. 
Inside the IC the wake potential is given by \cite{wlu06}, 
$(1+\beta)\frac{k_p^2 r_b^2 (\xi)}{4}-\frac{k_p^2 r^2}{4}$
where $k_p \equiv \omega_p/c$ and $\omega_p \equiv (n_e e^2/\epsilon_0 m_e)^{1/2}$ is the plasma frequency, with $n_e$ the plasma density. 
Note that these expressions are valid when the IC is created by long (negligible accelerating fields) or short pulse particle beams or lasers (large accelerating fields) and if there are large surface currents in the IC (as there is in the highly nonlinear channels). 
Therefore in all cases, the focusing force is $mc^2 k_p^2 r/2$ as was used in Ref. \cite{esar02}. 

In the focusing potential $\Phi(r)$, the Lorentz factor $\gamma$, the transverse radial position $r$ and the transverse radial momentum $p_r$ of an electron with an initial longitudinal momentum $p_0$, a maximum radius of oscillation $r_0$ and no azimuthal momentum, are given by $\gamma = \gamma_0 + r_0^2k_p^2\sin^2(\theta_r)/4$, $r = r_0\cos(\theta_r)$ and $p_r = K\sin(\theta_r)$ with $\gamma_0 = (1+p_0^2)^{1/2}$, $K = r_0k_p(\gamma_0/2)^{1/2}$ and $\theta_r = -Kct/r_0\gamma_0+\theta_{r0} = -\omega_\beta t+\theta_{r0}$, where $\theta_{r0}$ is the initial angle and $\omega_\beta=\omega_p/(2\gamma_0)^{1/2}$ is the betatron frequency.
Hereafter, the second order terms proportional to $\gamma_0^{-2}$ are neglected.
The electrons wiggling in the focusing potential generate a betatron radiation with a fundamental wavelength $\lambda_1=2\pi c/\omega_1$ with $\omega_1=4\gamma_0^2\omega_\beta/(2+K^2)$. 

The interaction between the electron beam and the radiation can lead to the amplification of the radiation.
In order to get micro-bunching, the spread in the radiation wavelength must be limited. 
In an ICL, the $K$ parameter depends on $r_0$ and $\gamma_0$ which can be different for each electron, so the radiation wavelength spread can be induced by both the beam energy spread and the $K$ spread.
A good approximation of the limiting spread can be found by assuming that $\Delta\lambda_1/\lambda_1<\rho$ should be satisfied, much in the same way as for FELs \cite{huan07}.
Knowing that $\lambda_1=2\pi c(2+K^2)(2\gamma_0)^{-3/2}/\omega_p$, we find that the energy spread and $K$ spread must then approximately satisfy the conditions given by Eq.(\ref{ICL_working_conditions}).

To further explore the optimal parameters for the ICL it is fundamental to determine the Pierce parameter.  
To start with, we analyze the bunching mechanism, which is a consequence of the energy exchange between the electrons and the radiation.
We first consider an electron propagating in the $z$ direction and a co-propagating EM wave.
This wave is polarized in the $x$ direction and characterized by its normalized vector potential:
\begin{equation}
A_x = A_1\cos(k_1z-\omega_1t+\Psi_1) \label{eq:def_Ax}
\end{equation}
where $A_1$ and $\Psi_1$ are respectively the wave amplitude and phase. 
We assume that the electron oscillates in the $(x,z)$ plan.
We then define $\phi$, the electron phase in the EM wave, and $\eta$, the relative electron energy as: 
\begin{eqnarray}
\phi & = & -\theta_r+k_1\overline{z}-\omega_1t \label{eq:def_phi} \\
\eta & = & \frac{\gamma_\eta-\gamma_0}{\gamma_0} \label{eq:def_eta} 
\end{eqnarray}
with $\gamma_\eta$ the electron Lorentz factor after its interaction with the wave and $\overline{z}$ is the longitudinal position of the electron averaged over one betatron oscillation. 
As shown in Appendix \ref{appA}, the interaction with the wave leads to the following equations of motion for the electron in the $(\phi,\eta)$ phase-space:
\begin{eqnarray}
\dot{\phi} & = & \frac{4+K^2}{8\gamma_0^2}\eta \label{eq:def_phi_p} \\
\dot{\eta} & = & \frac{A_1K[\hspace{-0.04cm}J\hspace{-0.09cm}J\hspace{-0.03cm}]}{2\gamma_0^2} \cos(\phi+\Psi_1) \label{eq:def_eta_p}
\end{eqnarray}
where $[\hspace{-0.04cm}J\hspace{-0.09cm}J\hspace{-0.03cm}]=J_0(K^2/(4+2K^2))-J_1(K^2/(4+2K^2))$, with $J_0$ and $J_1$ the Bessel functions.
Equation (\ref{eq:def_eta_p}) indicates that a beam of electrons is bunched by the EM wave at the phase $\phi=-\Psi_1+\pi/2+2m\pi$, with $m$ an integer, which leads to a bunching at the position $r=r_0\sin(k_1\overline{z}-\omega_1t+\Psi_1)$.
Therefore, due to the correlation between the radial and longitudinal position, the electron beam gets a continuous and oscillating shape after the bunching, with a period equal to $\lambda_1$. 
This is different from a conventional FEL, in which a succession of separated bunches is obtained.

Knowing the equations of motion, the amplification growth rate can be derived from the Vlasov and paraxial equations, as it has been described in \cite{huan07} for the conventional FEL case.
As explained in Appendix \ref{appB}, this method can be adapted to the ICL case by taking into account the equation (\ref{eq:def_phi_p}) and (\ref{eq:def_eta_p}).
The equivalent of the Pierce parameter $\rho_{\rm 1D}$ and the gain length of the radiation power $L_{GP}^{\rm 1D}$ in the 1D limit (radiation diffraction is neglected) for the ICL case is then given by:
\begin{eqnarray}
\rho_{\rm 1D}  & = &  \left[ \frac{I}{I_A} \frac{2(2+K^2)^2[\hspace{-0.04cm}J\hspace{-0.09cm}J\hspace{-0.03cm}]^2}{(4+K^2)^2\gamma_0} \right]^{1/3} \label{def_rho} \\
L_{GP}^{\rm 1D}  & = &  \frac{2(2+K^2)}{(4+K^2)\sqrt{3}\rho_{\rm 1D}}\frac{c}{\omega_\beta}
\end{eqnarray}
with $I$ the beam current and $I_A\sim 17$~kA the Alfv\`en current. 
We note that those results have been obtained assuming that $K\gg\rho^{1/2}$ and $\rho\ll1$.
Using $\rho\sim1$ may also lead to amplification, but the analytics have to be redone for this case. 

If the electrons have similar $\gamma_0$ and $K$ values, then the beam transverse size is limited to $2r_0=2K(2/\gamma_0)^{1/2}/k_p$. 
This is a major difference to conventional FEL, where this limitation is not present.
In an ICL, the radiation is emitted with a waist close to $r_0$, so the associated Rayleigh length is $Z_r\sim r_0^2k_1/2\ll L_{GP}^{\rm 1D}$.
As a result, the radiation diffraction can reduce or even stop the amplification and it should not be neglected.

As explained in Appendix \ref{appC}, taking into account the diffraction can lead to the following solution for the Pierce parameter and the power gain length:
\begin{eqnarray}
\rho    & = &  \rho_{\rm 1D}|\Gamma|^{\frac{1}{3}}  \label{def_pierce_multi_D} \\
L_{GP}  & = &  \frac{2+K^2}{8\gamma_0^2\Im(\mu)}\frac{c}{\omega_\beta}\label{def_LGP_multi_D}
\end{eqnarray}  
where $\Gamma$ and $\mu$ are given by:
\begin{eqnarray}
\Gamma  & = &  \int_{0}^{+\infty} -i\mu e^{i\mu\tilde{z}} B\!(\tilde{z},0) d\tilde{z} \label{def_Gamma_mu} \\
\mu     & = &  \frac{4+K^2}{8\gamma_0^2}\rho e^{i\left[ \frac{2\pi}{3} + \frac{1}{3}\arg(\Gamma)\right]}  \label{def_mu_multi_D}
\end{eqnarray}
with $B(z,r)$ the amplitude of a Gaussian beam characterized by its waist $W_0$, its wavelength $\lambda_1$ and $B(0,0)=1$.
$\rho$ and $L_{GP}$ correspond to the 2D or 3D solution, depending on if $B$ is the solution of respectively the 2D or 3D paraxial wave equation.
The solution of the coupled equations (\ref{def_pierce_multi_D}--\ref{def_mu_multi_D}) can be found iteratively: we start from the value of $\rho$ given by the 1D theory and $\Gamma=0$, then Eq.~(\ref{def_mu_multi_D}), (\ref{def_Gamma_mu}) and (\ref{def_pierce_multi_D}) can be solved iteratively until a converged solution is obtained.

In order to validate the theoretical conditions given in Eq.(\ref{ICL_working_conditions}), we have performed 2D simulations with the PIC code Osiris 2.0 \cite{fons02}. 
PIC codes are well suited to correctly and self-consistently model the radiation emission, diffraction, particle bunching and radiation amplification, as the full set of Maxwell's equations is solved.
As the typical IC size is much larger than the radiation wavelength $\lambda_1$, the IC formation is not self-consistently calculated in our simulation, allowing for a considerable reduction of the simulation size.
We initialize our simulations with a preformed field profile that matches the IC focusing fields.
A novel simulation technique that uses a Lorentz boosted frame \cite{jvay07, mart07} is used in order to considerably speed up the calculations, by performing simulations in the beam frame instead of the laboratory frame.
In this new frame, $\omega_\beta=\omega_1$, so the required number of time steps is reduced by a factor of $4\gamma_0^2/(2+K^2)$. 
For instance, a speed up of three orders of magnitude is obtained with $\gamma_0=50$ and $K=1$.
Moreover, running the ICL simulations in the beam frame prevents the numerical noise due to the numerical Cerenkov radiation \cite{godf74}.
The numerical noise can often perturb the bunching and artificially reduce or even stop the amplification. 
Perfectly matched layer (PML) absorbing boundary conditions \cite{jvay00} are used on the transverse side of the box, and periodic boundaries are used in the longitudinal direction. 
In the boosted frame, the box length was chosen between $2\lambda_r$ (for the shortest 3D simulations) and $40\lambda_r$ (for most of the 2D simulations). 
The box transverse size was typically equal to $40r_0$.
The longitudinal and transverse cell sizes used are typically $dz=dr=\lambda_r/50$.
We note that the self-consistent field amplitude in the simulation box is initially equal to 0, so the initial self-forces are neglected. This assumption is consistent with the fact that, in a FEL, the beam self-fields can be neglected as long as $\rho\ll1$ \cite{huan07}. 

\begin{figure}
\begin{center}
\includegraphics[width=0.38\textwidth]{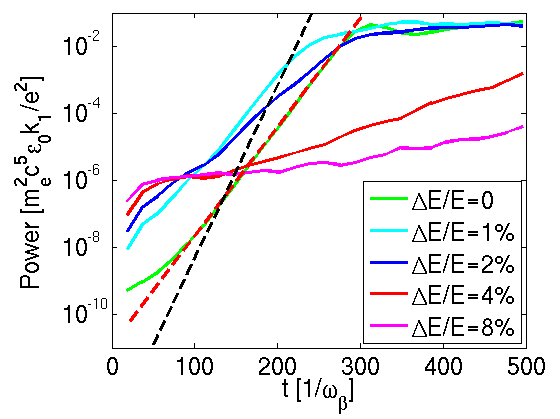}\hfill
\includegraphics[width=0.38\textwidth]{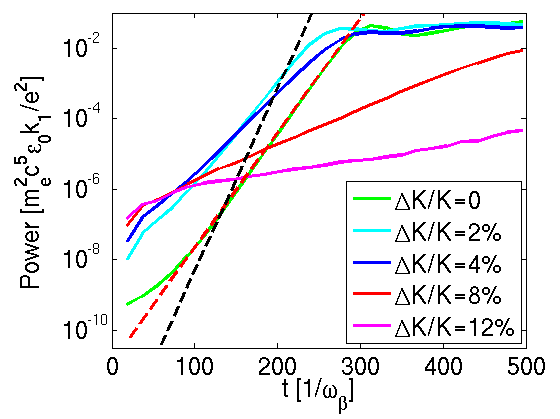}
\caption{Evolution of the radiation growth as a function of the energy spread (top) and $K$ spread (bottom). 
2D simulations with $\gamma_0=50$, $K=1$ and $I$=0.8~kA.
(Top) The green, light blue, dark blue, red and purple curves correspond to respectively $\Delta E/E=0$, $\Delta E/E=0.01$, $\Delta E/E=0.02$, $\Delta E/E=0.04$ and $\Delta E/E=0.08$.
(Bottom) The green, light blue, dark blue, red and purple curves correspond to respectively $\Delta K/K=0$, $\Delta K/K=0.02$, $\Delta K/K=0.04$, $\Delta K/K=0.08$ and $\Delta K/K=0.12$.
The dotted red and dotted black lines correspond to respectively the theoretical growth rate in the 1D limit and in 2D.
The $\gamma$ and $K$ spreads correspond to rms values. 
\label{growth_rate_K_E_spread} }
\end{center}
\end{figure}

In Fig.\ref{growth_rate_K_E_spread}, the simulation results for a beam characterized by $\gamma_0=50$, $K=1$ and a current $I$=0.8~kA injected in the IC fields are presented.
The beam parameters are chosen such that the computational costs of the simulations are reduced but the main physical features are captured. 
In the simulations, $\gamma$ and $K$ are initialized within a Gaussian distribution and the electrons are initialized with a random angle $\theta_{r0}$.
If $\Delta\gamma=\Delta K=0$, the Pierce parameter and power gain length determined in the 1D limit or in 2D are given by respectively $\rho_{\rm 1D}=0.082$, $L^{\rm 1D}_{GP}=8.4~c/\omega_\beta$, $\rho_{\rm 2D}=0.048$ and $L^{\rm 2D}_{GP}=13.4~c/\omega_\beta$.
The 1D and 2D theoretical growth rates are also represented in Fig.\ref{growth_rate_K_E_spread}.
We can observe a very good agreement between the 2D theoretical growth rate and the simulation results.
In the simulation, the initial noise produced by the macro-particles is amplified up to the saturation level.
This is reached when the particles are fully bunched.
However, with a high $\gamma$ or $K$ spread, the growth rate is reduced or even stopped. 
We observe that the change between a maximal and reduced growth rate matches the theoretical limits given by $\Delta K/K=0.072$ and $\Delta\gamma/\gamma=0.032$ with $\rho_{\rm 2D}=0.048$.

The condition $\Delta K/K\ll1$ can be parameterized by different complex configurations of the electron distribution in the transverse phase space.  
For example, in the 2D case, the electrons can be distributed over a ring in the transverse phase space.
This ring is parameterized by $r = r_0\cos(\theta_r)$ and $p_r = K\sin(\theta_r)$.
We propose more realistic distributions, with a spot shape instead of a ring shape.
In a first configuration, $K\sim1$ and $\Delta K/K\lesssim3\rho/2\ll1$ are used, so Eq.(\ref{ICL_working_conditions}) is satisfied, but the electrons are only distributed over a ring fraction, with an initial angle $\theta_{r0}$ that satisfies $|\theta_{r0}|<\theta_{r,max}$.
If $\theta_{r,max}\ll\pi$, the initial beam transverse size is much smaller than $r_0$ and the beam corresponds to an off-axis injected beam oscillating in the IC.
In that case, the beam shape in the transverse phase space is close to a spot with an initial transverse size and transverse momentum spread roughly equal to respectively $r_0\Delta K/K$ and $K\theta_{r,max}$.
In a second configuration, we choose $K\sim\rho^{1/2}\ll1$ and $\Delta K/K\sim1$, which still satisfies the $K$ spread condition in Eq.(\ref{ICL_working_conditions}). 
In that case, as $\Delta K/K\sim1$, the maximum radial momentum $p_{rm}$ of a given electron roughly satisfies $K-\Delta K\lesssim p_{rm}\lesssim K+\Delta K$ so $|p_{rm}|\lesssim 2K$.
We also have $|r_{0m}|\lesssim 2r_0$ with $r_{0m}$ the maximum radial position of a given electron.
Therefore, the spread around the ring is such that the beam distribution in the transverse phase space becomes a spot.
As $r_0\propto K$, using $K\ll1$ corresponds to a narrow on-axis injected beam.

\begin{figure}
\begin{center}
\includegraphics[width=0.38\textwidth]{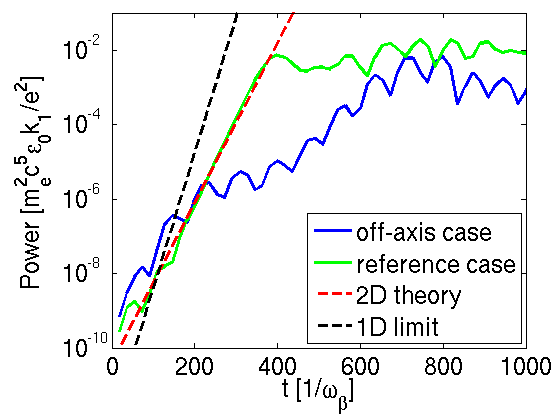}\hfill
\includegraphics[width=0.38\textwidth]{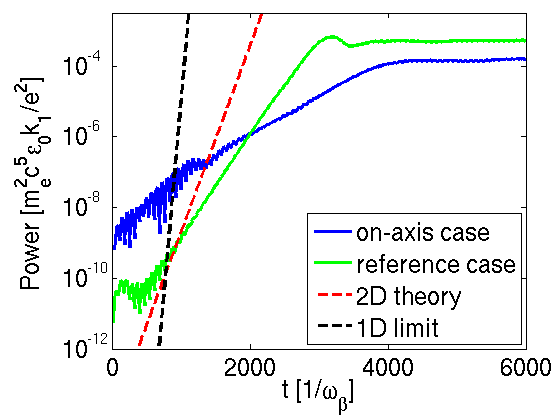}
\caption{(Color online) (Top) In blue: radiated power for an off-axis beam, initialized with $|\theta_{r0}|<\pi/16$, $K=1$, $\Delta K/K=0.01$ and $\Delta\gamma/\gamma=0.005$. 
In green: radiated power in the reference case, with $|\theta_{r0}|<\pi$ and $\Delta K/K=\Delta\gamma/\gamma=0$. 
(Bottom) In blue: radiated power for an on-axis beam, which is initialized with $|\theta_{r0}|<\pi$, $K=0.1$, $\Delta K/K=0.3$ and $\Delta\gamma/\gamma=0.002$. 
In green: radiated power in the reference case, with $\Delta K/K=\Delta\gamma/\gamma=0$.
The dotted red and dotted black lines correspond to respectively the theoretical growth rate in the 1D limit and in 2D.
\label{growth_rate_realist_case} }
\end{center}
\end{figure}

The two configurations are highlighted by 2D simulations.
In the first case, an off-axis beam with $\gamma_0=50$, $K=1$ and $I$=0.27~kA is injected with $|\theta_{r0}|<\pi/16$. 
The corresponding Pierce parameter is $\rho_{\rm 2D}=0.031$ and the beam is initialized with $\Delta K/K=$0.01 and $\Delta\gamma/\gamma=0.005$.
In the second case, an on-axis beam with $\gamma_0=50$, $K=0.1$ and $I$=42~A is injected with $|\theta_{r0}|<\pi$.
The corresponding Pierce parameter is $\rho_{\rm 2D}=6.4\times10^{-3}$ and the beam is initialized with $\Delta K/K=0.3$ and $\Delta\gamma/\gamma=0.002$.
Reference simulations have been performed for both cases, using $|\theta_{r0}|<\pi$ and $\Delta K/K=\Delta\gamma/\gamma=0$.
The evolution of the amplified radiation power for those different simulations is presented in Fig.\ref{growth_rate_realist_case}.
In both cases, we observe that the use of more realistic beams, with a finite spot in the transverse phase space and an energy spread, can still lead to exponential radiation amplification, even if the growth rate and final power are lower than in the reference simulations, for the idealized scenarios.
The discrepancy between the 2D theoretical growth rate and the idealized simulation result in the on-axis case is due to the use of $K=1.25\rho^{1/2}$ whereas our theoretical model is valid in the limit $K\gg\rho^{1/2}$.

\begin{figure}
\begin{center}
\includegraphics[width=0.38\textwidth]{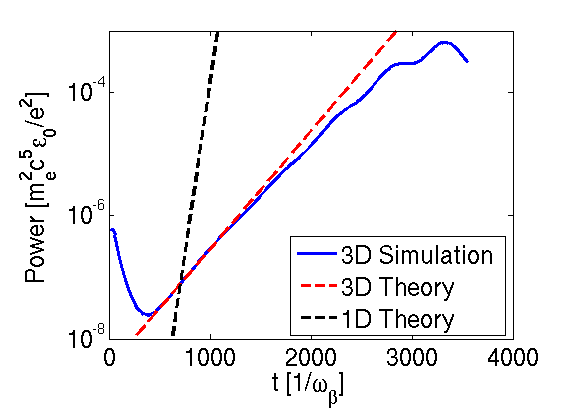}\hfill
\includegraphics[width=0.48\textwidth]{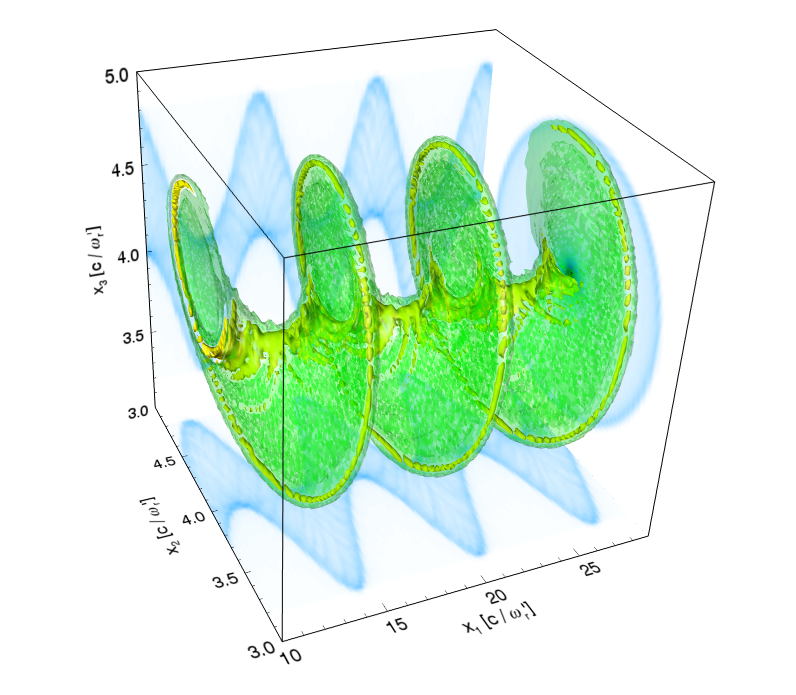}
\caption{(Color online) (Top) Radiation growth with $\gamma_0=50$, $K=1$, $I$=8~A in a 3D simulation (blue) and given by the 1D theory (dotted black) and 3D theory (dotted red). (Bottom) Shape of the electron beam at saturation in the 3D simulation with $\gamma_0=50$, $K=1$, $I$=0.8~kA: a helical bunching is observed (iso-surface of the electron density).
\label{simu_3D} }
\end{center}
\end{figure}

We have also performed 3D simulations to confirm the 3D theoretical results.
The electrons are initialized with a radial momentum but no azimuthal momentum.
The results are shown in Fig.\ref{simu_3D}.
In the simulation with $\gamma_0=50$, $K=1$ and $I$=8~A, the corresponding Pierce parameter and power gain length obtained in the 1D limit or in 3D are given by respectively $\rho_{\rm 1D}=0.018$, $L^{\rm 1D}_{GP}=39~c/\omega_\beta$, $\rho_{\rm 3D}=2.7\times10^{-3}$ and $L^{\rm 3D}_{GP}=226~c/\omega_\beta$.
A good agreement between simulation and theory is found in 3D.
Since the initial noise in the simulation is too low to start the amplification mechanism in the 3D simulations, we have injected a seed in the IC.
The seed wavelength is $\lambda_1$, like the expected amplified radiation. 
As the seed diffracts, most of its energy gets out from the simulation box from the side.
This explains the power dip at the beginning of the simulation at $t\sim 500~\omega_\beta^{-1}$.
The amplification is initiated and the saturation level is reached at the end of the simulation. 
At saturation, the bunch shape is helical. 
This result is consistent with a circularly polarized seed. 

Our results show that an amplification of several orders of magnitude of the radiated power can be achieved, even if the Rayleigh length of the generated radiation is shorter than the gain length in the typical ICL configurations.
The diffraction is responsible for the growth rate reduction.
The gain length is 1.6 times larger than the 1D limit in the 2D case presented in Fig.\ref{growth_rate_K_E_spread}, and 5.8 times in this 3D case.
We have also confirmed in our ab initio simulations that the amplified radiation wavelength and the oscillating shape period of the bunching are $\lambda_1$, matching the theory.
Odd harmonics have also been observed.
Yet, as expected with $K=1$, their amplitudes are much smaller than the fundamental harmonic amplitude.    
This demonstrates that PIC simulations in the beam frame might be an efficient tool to study the self-consistent dynamics of harmonics and their feedback on the growth rate in scenarios where $K>1$ in an ICL or in FELs. 

It is important to note that the accelerating field, present in a typical IC can affect the amplification process, as the electron energy will change in time. 
This effect can be reduced, for instance, by injecting the electron beam close to the center (longitudinally) of the bubble in a wakefield, where the accelerating field is zero.
However, the accelerating field effect study is beyond the scope of this paper and will be considered in future publications.  

Numerical applications of our analytical results show that the most stringent condition will be to inject an electron beam with a very low emittance and at a precise radius in the bubble. 
For instance, if we consider a 500~TW laser pulse driving a wakefield in a plasma with a density $n_e=5\times10^{17}$~cm$^{-3}$, the injection of a 250~MeV, 10~kA beam with $K=1$ produce an X-ray beam with a wavelength of 4.6~nm. 
The associated gain length is $L^{\rm 3D}_{GP}=5.6$~mm, so the radiated power can be multiplied by 1000 after 3.8~cm. 
To get this amplification, the electron beam should have a relative energy spread lower than 1.7\% and be injected at 0.48~microns off axis with a normalized transverse emittance of 0.01~mm.mrad. 
Even if the emittance value is still far from the best values obtained in a wakefield accelerator, optimization or mix of new injection scheme in a wakefield accelerator, such as optical \cite{faur07,davo09}, ionization \cite{apak10,mcgu10} or magnetic \cite{viei11} injection, might help to improve emittance and control the off axis injection. 

In this paper, we have described the required conditions on the electron beam quality in order to observe ion channel lasing.
The Pierce parameter and amplification growth rate have been determined analytically taking into account the effects of diffraction. 
It is shown that it is not necessary to use a guiding structure for the radiation. 
2D and 3D PIC simulations, which are the first fully relativistic electromagnetic 3D simulations of ICL, have confirmed our analytical findings, illustrating the possibility of achieving high-gain radiation amplification in ICL. 
These results pave the way for the generation of high brilliance coherent radiation in compact plasma structures.

\appendix

\section{Electron motion equations in the $(\phi,\eta)$ phase-space}
\label{appA}

In this Appendix, we show that the interaction between an electron following a betatron motion and an EM defined by Eq. (\ref{eq:def_Ax}) leads to the motion equations in the $(\phi,\eta)$ phase-space given by Eq. (\ref{eq:def_phi_p}) and (\ref{eq:def_eta_p}).  

In an ICL, $K$ is a function of $\gamma$. 
Therefore, we first need to determine how $K$ evolves, as well as $\gamma_0$, $p_0$ or $r_0$, when the EM wave exchanges energy with the electron.
This result is first presented and the description of the bunching process, leading to Eq.~(\ref{eq:def_phi_p}) and (\ref{eq:def_eta_p}), is addressed in the second part of this Apendix.

\subsection{Influence of an EM wave on the betatron oscillation parameters}

As mentioned, $K$ can evolve and is now a function of time. 
We defined $K_0$ and the time-dependent longitudinal momentum $p_z$ and maximum radius $r_m$ such that $K_0=K(t=0)$, $p_z(t=0)=p_0$ and $r_m(t=0)=r_0$.
We still consider that $\gamma_0=(1+p_0^2)^{1/2}$.
In the following, we use the notation:
\begin{equation}
\dot{X} = \frac{dX}{d(\omega_1t)}
\end{equation}
In the presence of an EM wave defined by Eq.~(\ref{eq:def_Ax}), the energy and momentum change of an electron following betatron motion in the $(x,z)$ plan is given by:     
\begin{eqnarray}
\dot{\gamma}  & = &  \beta_r\alpha - \beta_r\frac{K(2+K^2)}{4\gamma^2}\cos(\theta_r)  \label{dg_dt} \\
\dot{p_r}     & = &  (1-\beta_z)\alpha -  \frac{K(2+K^2)}{4\gamma^2}\cos(\theta_r)   \label{dpr_dt} \\
\dot{p_z}     & = &  \beta_r\alpha   \label{dpz_dt}
\end{eqnarray}
where $\alpha = A_1\sin(k_1z-\omega_1t+\Psi_1)$ and $\beta_r$ and $\beta_z$ are the normalized transverse and longitudinal electron velocities. 
In the above equation, we have used the following identity to define the ion-channel focusing field, which is normalized to $m_ec\omega_1/e$: 
\begin{equation}
E_x = \frac{rk_p}{2}\frac{\omega_p}{\omega_1} = \frac{K(2+K^2)}{4\gamma^2}\cos(\theta_r)
\end{equation}

By using $r=r_m\cos(\theta_r)$, $p_r=K\sin(\theta_r)$ and $K=r_mk_p(\gamma/2)^{1/2}$, we can show that:
\begin{eqnarray}
K      & = & \sqrt{ p_r^2 + \frac{\gamma}{2}r^2k_p^2 }  \\
r_mk_p & = & \sqrt{ \frac{2}{\gamma }p_r^2 + r^2k_p^2 }
\end{eqnarray}
Therefore:
\begin{eqnarray}
\dot{K}       & = &  \frac{ 4p_r\dot{p_r} + \dot{\gamma}r^2k_p^2 + 2\gamma r\dot{r}k_p^2 }{4K}  \\
\dot{r_m}k_p  & = &  \frac{ -\dot{\gamma}p_r^2 + 2\gamma p_r\dot{p_r} + \gamma^2r\dot{r}k_p^2 }{r_mk_p\gamma^2} 
\end{eqnarray}
As $\dot{K}=0$ and $\dot{r_m}=0$ when $\alpha=0$, we can simplify the equation and get: 
\begin{eqnarray}
\dot{K}             & = &  \frac{ 1+K^2 }{ 2\gamma^2 } \alpha\sin(\theta_r)  \\
\dot{r_m}k_p   & = &   \frac{ \alpha\sin(\theta_r) }{ \gamma^2\sqrt{2\gamma} } 
\end{eqnarray}

We now consider the average of the derivatives over one betatron period, and we assume that the change of $\gamma$, $K$ and $r_m$ is small during one betatron period ($\dot{\gamma} \ll \omega_\beta\gamma$). 
The averaged derivatives are then given by:
\begin{eqnarray}
\dot{\overline{\gamma}}  & = &  \overline{\beta_r\alpha}   \label{eq:dgm_dt} \\           
\dot{\overline{K}}       & = &  \frac{ 1+\overline{K}^2 }{ 2\overline{K}\overline{\gamma} } \dot{\overline{\gamma}}  \\
\dot{\overline{r_m}}k_p  & = &   \frac{ 1 }{ \overline{K}\overline{\gamma}\sqrt{2\overline{\gamma}} } \dot{\overline{\gamma}} 
\end{eqnarray}
We introduce the parameter $\nu$ defined as:
\begin{equation}
\nu = \frac{\dot{\overline{\gamma}}}{\overline{\gamma}}
\end{equation}
Then, we obtain:
\begin{eqnarray}  
\frac{\dot{\overline{K}}}{\overline{K}}           & = &  \frac{ 1+\overline{K}^2 }{ 2\overline{K}^2 } \nu  \\
\frac{\dot{\overline{r_m}}}{\overline{r_m}}    & = &   \frac{ 1 }{ 2\overline{K}^2 } \nu 
\end{eqnarray}
We also assume that all parameter evolutions are small during the whole interaction (e.g. $\gamma(t)-\gamma_0 \ll \gamma_0$ for all $t$).   
This leads to:
\begin{eqnarray}  
\overline{\gamma}  & = &  \gamma_0 \left( 1 + \frac{1}{\gamma_0}\int_0^t \dot{\overline{\gamma(\tau)}}d\tau  \right) \nonumber \\  
                   & = &  \gamma_0 \left( 1 + \int_0^t \nu d\tau  \right) \\
\overline{K}       & = &  K_0  \left( 1 + \frac{ 1+\overline{K}^2 }{ 2\overline{K}^2 } \int_0^t \nu d\tau \right)  \\
\overline{r_m}     & = &  r_0  \left( 1 +  \frac{ 1 }{ 2\overline{K}^2 } \int_0^t \nu d\tau \right) 
\end{eqnarray}
We introduce $\eta=\int_0^t \nu d\tau$.
According to our last assumption we have $\eta \ll 1$ and $K_0~\gg~\eta^{1/2}$.
By neglecting the terms proportional to $\eta^2$, we then obtain:
\begin{eqnarray}  
\eta            & = & \frac{\gamma_\eta-\gamma_0}{\gamma_0} \\
K_\eta       & = & K_0  \left( 1 + \frac{ 1+K_0^2 }{ 2K_0^2 } \eta \right)  \label{eq_def_K_eta} \\
r_\eta        & = & r_0  \left( 1 +  \frac{ 1 }{ 2K_0^2 } \eta \right) 
\end{eqnarray}
where we rename $\overline{\gamma}$, $\overline{K}$ and $\overline{r_m}$ by respectively $\gamma_\eta$, $K_\eta$ and $r_\eta$ for convenience. 

Hereafter, all the items with $\eta$ as subscript are function of $\gamma_\eta$, $K_\eta$ or $r_\eta$, and if $\eta$ is not mentioned, it means that the value is taken at $\eta=0$. 
For example, $\omega_{\beta\eta}=\omega_p/(2\gamma_\eta)^{1/2}$, and from this equation we find:
\begin{equation}
\omega_{\beta\eta} = \omega_\beta \left( 1-\frac{\eta}{2} \right)
\end{equation}
We also deduce from $\omega_\beta/\omega_1 =  (2+K_0^2)/(4\gamma_0^2)$ that:
\begin{equation}
\omega_{1\eta} = \omega_1 \left( 1+\frac{4+K_0^2}{4+2K_0^2}\eta \right)
\label{def_omega_1eta}
\end{equation}

\subsection{Electron motion in the presence of an EM wave}

We can now rewrite Eq.~(\ref{eq:dgm_dt}) as: 
\begin{equation}
\dot{\overline{\gamma}} = \dot{\gamma_\eta}=  \frac{A_1K_\eta}{\gamma_\eta} \overline{ \sin(\theta_r) \sin( k_1z-\omega_1t+\Psi_1)  } 
\end{equation}
Here, only the dominant term is relevant and the terms proportional to $\eta$ can be neglected.
This is also true for the terms in the phase $k_1z-\omega_1t+\Psi_1$.
Thus, we determine $k_1z$ while neglecting the terms proportional to $\eta$:
\begin{eqnarray}
k_1z & =    &  k_1\int_0^tc\beta_zdt + z_0k_1  \nonumber  \\
     & \sim &  \omega_1\int_0^t\left( 1-\frac{2+K_0^2}{4\gamma_0^2}+\frac{K_0^2}{4\gamma_0^2}\cos(2\theta_r) \right)dt + z_0k_1 \nonumber \\
     & \sim &  \omega_1t + \theta_r - \theta_{r0} - \frac{K_0^2}{4+2K_0^2}(\sin(2\theta_{r})-\sin(2\theta_{r0})) \nonumber \\
     &      &  + z_0k_1  \nonumber \\
     & \sim &  k_1\overline{z} - \frac{K_0^2}{4+2K_0^2}\sin(2\theta_{r})
\label{def_k_1z}
\end{eqnarray}
where $z_0$ and $\theta_{r0}$ are the initial position and phase.
We define a new phase $\phi$ as
\begin{equation}
\phi = -\theta_r+k_1\overline{z}-\omega_1t 
\label{def_phi}
\end{equation}
By using Eq.(\ref{def_k_1z}), we can note that $\phi$ is a constant of time when we neglect the terms proportional to $\eta$, so $\overline{\phi}=\phi$.
We then find that:
\begin{widetext}
\begin{eqnarray}
\dot{\gamma_\eta} & = & \frac{A_1K_0}{\gamma_0} \overline{ \sin(\theta_r) \sin\left( \phi + \Psi_1 + \theta_r - \frac{K_0^2}{4+2K_0^2}\sin(2\theta_{r})\right)  }    \nonumber \\
\dot{\gamma_\eta} & = & \frac{A_1K_0}{2\gamma_0} \cos(\phi+\Psi_1) \left[ \overline{\cos\left( -\frac{K_0^2}{4+2K_0^2}\sin(2\theta_{r}) \right)} - \overline{\cos\left( 2\theta_r-\frac{K_0^2}{4+2K_0^2}\sin(2\theta_{r}) \right)} \right]     \nonumber \\
                                  &     & \hspace{-0.23cm} + \frac{A_1K_0}{2\gamma_0} \sin(\phi+\Psi_1) \left[ \overline{\sin\left( -\frac{K_0^2}{4+2K_0^2}\sin(2\theta_{r}) \right)} + \overline{\sin\left( 2\theta_r-\frac{K_0^2}{4+2K_0^2}\sin(2\theta_{r}) \right)} \right]
\end{eqnarray}
\end{widetext}
By using the following identity 
\begin{eqnarray}  
\overline{\sin\left( 2\theta_r - \frac{K_0^2}{4+2K_0^2}\sin(2\theta_r) \right)} & = & \nonumber \\
\overline{\sin\left( \frac{K_0^2}{4+2K_0^2}\sin(2\theta_r) \right)} & = & 0
\end{eqnarray}
\begin{equation} 
\overline{\cos\left( \frac{K_0^2}{4+2K_0^2}\sin(2\theta_r) \right)}  =  J_0\left(\frac{K_ 0 ^2}{4+2K_ 0 ^2}\right)
\end{equation}
\begin{eqnarray} 
& \displaystyle \overline{\cos\left( 2\theta_r - \frac{K_0^2}{4+2K_0^2}\sin(2\theta_r) \right)}  = & \nonumber \\
& \displaystyle J_1\left(\frac{K_ 0 ^2}{4+2K_ 0 ^2}\right) &   
\end{eqnarray}
where $J_0$ and $J_1$ are the Bessel functions, we can then write:
\begin{equation}
\dot{\gamma_\eta} =   \frac{A_1K_0[\hspace{-0.04cm}J\hspace{-0.09cm}J\hspace{-0.03cm}]}{2\gamma_0} \cos(\phi+\Psi_1)
\end{equation}
where $[\hspace{-0.04cm}J\hspace{-0.09cm}J\hspace{-0.03cm}]=J_0(K_0^2/(4+2K_0^2))-J_1(K_0^2/(4+2K_0^2))$.

To get the derivative of $\phi$ with time, we need to rewrite Eq.(\ref{def_phi}) without neglecting the terms proportional to $\eta$. 
Provided that:
\begin{eqnarray}
\frac{d(k_1\overline{z})}{d(\omega_1t)}    & = &  \overline{\beta_z} =  1-\frac{\omega_{\beta\eta}}{\omega_{1\eta}} \\
\frac{\omega_{\beta\eta}}{\omega_{1\eta}}  & = &  -\dot{\theta}_r - \frac{4+K_0^2}{8\gamma_0^2}\eta 
\end{eqnarray}
we eventually obtain the equation of motion in the $(\phi,\eta)$ phase space:
\begin{eqnarray}  
\dot{\phi} & = &  \frac{4+K_0^2}{8\gamma_0^2}\eta  \label{def_phi_p} \\
\dot{\eta} & = & \frac{A_1K_0[\hspace{-0.04cm}J\hspace{-0.09cm}J\hspace{-0.03cm}]}{2\gamma_0^2} \cos(\phi+\Psi_1)   \label{def_eta_p}
\end{eqnarray}

\section{Coupling between the Maxwell and motion equations}
\label{appB}

In this Appendix, we follow the method used in Ref.~\cite{huan07} to calculate the 1D and ideal growth rate for FEL.
Here, this method has been adapted to the ICL context.

To start with, we consider the presence of an EM wave polarized along the $x$ direction, propagating along the $z$ direction, and characterized by its normalized vector potential $A_x=A_0(x,y,\xi,\tau)$, where $\xi=k_1z-\omega_1t$ and $\tau=\omega_1t$. 
We introduce the wave amplitude $A_\nu(x,y,\tau)$ in the frequency domain through:
\begin{equation}
A_0(x,y,\xi,\tau) = \frac{1}{2} \int_{0}^{+\infty} A_\nu(x,y,\tau)e^{i\nu\xi}d\nu \ + \ c.c.
\label{def_A0_fft}
\end{equation}
where $c.c.$ is the complex conjugate.
The Maxwell equations for $A_\nu$ give:
\begin{equation}
\displaystyle \left( \frac{\partial^2}{\partial\tau^2} - 2\frac{\partial^2}{\partial\xi\partial\tau} - \boldsymbol{\nabla}_\bot^2 \right) (A_\nu e^{i\nu\xi}) = \frac{4\pi e\omega_1}{I_A} j_\nu e^{i\nu\xi}   
\end{equation}
\begin{equation}
\displaystyle  j_\nu = \frac{1}{\pi} \int_{-\infty}^{+\infty} j_x(x,y,\xi,\tau) e^{-i\nu\xi}d\xi
\end{equation}
where $\boldsymbol{\nabla}_\bot^2$ is the transverse Laplacian normalized to $k_1^2$.
$j_x$ is the transverse current density along the $x$ direction, and it is normalized to $e\omega_1k_1^2$.
$I_A=ec/r_e$ is the Alfv\`en current, with $r_e=e^2/(4\pi\epsilon_0m_ec^2)$ the classical electron radius.
By using the slowly varying envelope approximation $(|\partial^2 A_\nu/\partial\tau^2| \ll 2\nu|\partial A_\nu/\partial\tau|)$, we get:
\begin{equation}
\left(  2i\nu \frac{\partial}{\partial\tau} + \boldsymbol{\nabla}_\bot^2 \right) A_\nu = - \frac{4\pi e\omega_1}{I_A} j_\nu 
\label{eq_wave_slow_vary_env}
\end{equation}

\subsection{Calculation of the transverse current}

The normalized transverse current density $j_{xn}$ of the particle $n$, which follows betatron motion in the $(x,z)$ plan is given by:
\begin{eqnarray}
j_{xn}(x,\xi,\tau) & = & - \frac{K_n}{\gamma_n}\sin(\theta_{rn})\delta(k_1x-k_1x_n)\delta(\xi-\xi_n)  \nonumber \\
                   & = & - \frac{K_n}{2i\gamma_n}(e^{i\theta_{rn}}-e^{-i\theta_{rn}})\delta(k_1x-k_1x_n) \nonumber \\
                   &   & \times\delta(\xi-\xi_n)
\end{eqnarray}
where $K_n$, $\gamma_n$ and  $\theta_{rn}$ are the parameters of the electron at the time $\tau$, $\delta$ is the Dirac function, and $x_n$ and $\xi_n$ are respectively the transverse position of the electron and its position over the $\xi$ direction at the time $\tau$.
$x_n$ and $\xi_n$ are given by:
\begin{eqnarray}
x_n   & = &  r_{m,n}\cos(\theta_{rn})       \label{def_x_n}  \\
\xi_n & = &  \phi_n + \theta_{rn} -\frac{K_n^2}{4+2K_n^2}\sin(2\theta_{rn})   \label{def_xi_n}   
\end{eqnarray}
where $r_{m,n}$ is the maximum radius of oscillation of the electron $n$ at time $\tau$.
Eq.(\ref{def_xi_n}) is obtained by using Eq.(\ref{def_phi}).
$j_{\nu n}$ is then given by:
\begin{eqnarray}
j_{\nu n}(x,\tau)  & = & - \frac{K_n}{2i\pi\gamma_n} e^{-i(\Delta\nu\theta_{rn}+\nu\phi_n)}  \nonumber  \\
                   &   & \times \delta[k_1x-k_1r_{m,n}\cos(\theta_{rn})]   \nonumber  \\
                   &   &  \times  e^{i\nu\frac{K_n^2}{4+2K_n^2}\sin(2\theta_{rn})}  \left( 1-e^{-2i\theta_{rn}} \right) 
\end{eqnarray}
Where $\Delta\nu=\nu-1$. 
We define the function $G(x,\nu,K,\gamma,\theta_r)$ as:
\begin{eqnarray}
G(x,\nu,K,\gamma,\theta_r) & = & \delta[k_1x-k_1r_{m}\cos(\theta_{r})] \nonumber \\
                           &   & \times  e^{i\nu\frac{K^2}{4+2K^2}\sin(2\theta_{r})} \nonumber \\
                           &   & \times   \left( 1-e^{-2i\theta_{r}} \right)   
\end{eqnarray}
where $r_m$ is a function of $\gamma$ and $K$, since we have the following identity:
\begin{equation}
k_1r_m  = \frac{4\gamma K}{2+K^2}
\end{equation}
The current created by the electron $n$ is then given by:
\begin{equation}
j_{\nu n}(x,\tau)  =  - \frac{K_n}{2i\pi\gamma_n} e^{-i(\Delta\nu\theta_{rn}+\nu\phi_n)} G(x,\nu,K_n,\gamma_n,\theta_{rn}) 
\end{equation}

The electron distribution at the time $\tau$ in the phase space can be parameterized by the 4 parameters $\phi$, $\eta$, $K$ and $\theta_r$.
Therefore, the distribution function $F$ is given by:
\begin{eqnarray}
& \displaystyle F(\phi,\eta,K,\theta_r,\tau) = \frac{2\pi e\omega_1}{I} \times  \nonumber \\
& \displaystyle \sum_{n=1}^{n=N} \delta(\phi-\phi_n) \delta(\eta-\eta_n) \delta(K-K_n) \delta(\theta_r-\theta_{rn}) \ \ 
\end{eqnarray}
where $I$ is the longitudinal beam current (absolute value so $I>0$), $N$ is the number of electrons and $2\pi e\omega_1/I$ is a normalizing factor. 
The total current $j_{\nu}(x,\tau)$ is then given by:
\begin{eqnarray}
j_{\nu}(x,\tau)  & = &  - \int  \frac{IK}{2i\pi\gamma e\omega_1} e^{-i(\Delta\nu\theta_{r}+\nu\phi)}   \nonumber \\
                 &   &  \hspace{0.75cm}    \times  G(x,\nu,K,\gamma,\theta_r) F(\phi,\eta,K,\theta_r,\tau) \nonumber \\
                 &   &  \hspace{0.75cm}    \times  d\phi d\eta dK \frac{d\theta_r}{2\pi}   \label{j_nu_integral_form}
\end{eqnarray}
where, according to the definition of $\eta$, we have $\gamma=\gamma_0(1+\eta)$.

The normalizing factor of $F$ has been chosen so that if we consider a beam distribution with the parameters $\phi$, $\eta$, $K$ and $\theta_r$ which are not correlated, then $F$ can be written as:
\begin{equation}
F = D_1(\phi,\tau) D_2(\eta,\tau) D_3(K,\tau) D_4(\theta_r,\tau)
\label{def_F_cas_ideal_part_1}
\end{equation}
where
\begin{eqnarray}
\int D_1 d\phi      & = &  Lk_1 \label{def_F_cas_ideal_part_2_D1}  \\
\int D_2 d\eta      & = &  1 \label{def_F_cas_ideal_part_2_D2}  \\
\int D_3 dK         & = &  1 \label{def_F_cas_ideal_part_2_D3}  \\
\int D_4 d\theta_r  & = &  2\pi \label{def_F_cas_ideal_part_2_D4}
\end{eqnarray}
with $L=Nec/I$ the beam length.

\subsection{1D approximation}

We now consider that the EM wave is a plane wave, so the term $\boldsymbol{\nabla}_\bot^2$ and the transverse position can be neglected. 
Eq.(\ref{eq_wave_slow_vary_env}) then becomes:
\begin{equation}
\frac{\partial A_\nu(\tau)}{\partial\tau}   = -\frac{2\pi e\omega_1}{i\nu I_A} j_\nu(\tau)
\label{eq:paraxial_eq_1D}
\end{equation}
where $j_\nu(\tau)$ is the current averaged over the beam transverse size and is given by:
\begin{eqnarray}
j_{\nu}(\tau)  & = &  - \frac{1}{Sk_1^2} \int  \frac{IK}{2i\pi\gamma e\omega_1} e^{-i(\Delta\nu\theta_{r}+\nu\phi)}  \nonumber \\
               &   &  \hspace{1.3cm}    \times  G_2(\nu,K,\gamma,\theta_r) F(\phi,\eta,K,\theta_r,\tau) \ \ \ \ \  \nonumber \\
               &   &  \hspace{1.3cm}    \times  d\phi d\eta dK \frac{d\theta_r}{2\pi}   \label{eq_j_nu_tau_mean_transverse}  
\end{eqnarray}
\begin{equation}
G_2(\nu,K,\gamma,\theta_r)  =  e^{i\nu\frac{K^2}{4+2K^2}\sin(2\theta_{r})}   \left( 1-e^{-2i\theta_{r}} \right) 
\end{equation}
with $S=\pi r_0^2$ the beam transverse size.
Since only the radiations with a wavelength close to $\lambda_1$ are generated and amplified, we assume that $\Delta\nu\ll1$, so the term $\Delta\nu\theta_{rn}+\nu\phi_n$ evolves slowly and can be considered as constant over one betatron period.
We also get $\overline{G_2}=[\hspace{-0.04cm}J\hspace{-0.09cm}J\hspace{-0.03cm}]$.
The average over one betatron period of the current is then:
\begin{eqnarray}
\overline{j_{\nu}}(\tau) & = & -\frac{IK_0[\hspace{-0.04cm}J\hspace{-0.09cm}J\hspace{-0.03cm}]}{2i\pi^2\gamma_0r_0^2k_1^2e\omega_1}  \times \nonumber \\
&& \int  e^{-i(\Delta\nu\theta_{r}+\nu\phi)} F(\phi,\eta,K,\theta_r,\tau) \nonumber \\
&&  \hspace{0.35cm} \times  d\phi d\eta dK \frac{d\theta_r}{2\pi}
\end{eqnarray}
where we have also considered a small energy and $K$ spread, so $K\sim K_0$ and $\gamma\sim\gamma_0$.
We finally get:
\begin{eqnarray}
\frac{\partial (\nu A_\nu)}{\partial\tau} &  = & - \frac{ 2K_0[\hspace{-0.04cm}J\hspace{-0.09cm}J\hspace{-0.03cm}]}{\gamma_0r_0^2k_1^2}\frac{I}{I_A}  e^{i\Delta\nu\frac{\omega_\beta}{\omega_1}\tau}  \times \nonumber \\
&&  \int  \frac{e^{-i\nu\phi}}{2\pi} F(\phi,\eta,K,\theta_r,\tau) d\phi d\eta dK \frac{d\theta_r}{2\pi} \hspace{1.0cm} 
\end{eqnarray}
where we have also assumed that $\exp[-i\Delta\nu\theta_r]\sim \exp[i\Delta\nu(\omega_\beta/\omega_1)\tau]$.
Indeed, as we have the three identities $\dot{\theta_r}=-\omega_{\beta\eta}/\omega_1$, $\omega_{\beta\eta}\sim\omega_\beta$ and $\Delta\nu\ll1$, then the difference between $\Delta\nu\theta_r$ and $-\Delta\nu(\tau\omega_{\beta}/\omega_1+\theta_{r0})$ is small, even if $\tau\gg1$. 
Moreover, according to Eq.(\ref{def_x_n},\ref{def_xi_n}), it is possible to choose $\theta_{r0}$ so that $\theta_{r0}\in[0,2\pi]$, so $\Delta\nu\theta_{r0}\ll1$.

The Vlasov equation is defined by $\dot{F}=0$. 
We thus get:
\begin{equation}
\frac{\partial F}{\partial \tau} + \dot{\phi}\frac{\partial F}{\partial \phi} + \dot{\eta}\frac{\partial F}{\partial \eta} + \dot{K}\frac{\partial F}{\partial K} + \dot{\theta_r}\frac{\partial F}{\partial \theta_r} = 0
\end{equation}
Moreover, with an EM wave described by Eq.(\ref{def_A0_fft}), Eq.(\ref{def_eta_p}) becomes:
\begin{equation}
\dot{\eta}  =  \frac{K_0[\hspace{-0.04cm}J\hspace{-0.09cm}J\hspace{-0.03cm}]}{4\gamma_0^2} \int_{0}^{+\infty} \nu A_\nu e^{i(\Delta\nu\theta_r+\nu\phi)} d\nu \ + \ c.c.
\end{equation}
where the dependence of $[\hspace{-0.04cm}J\hspace{-0.09cm}J\hspace{-0.03cm}]$ as a function of $\nu$ has been neglected. 

We introduce the following scaled variables to simplify our equation:
\begin{eqnarray}
\hat{\eta} & = & \frac{\eta}{\rho} \\
\hat{\tau} & = & \frac{4+K_0^2}{8\gamma_0^2}\rho\tau  \label{def_hat_tau}  \\
\Delta\hat{\nu} & = & \frac{8\gamma_0^2}{\rho(4+K_0^2)} \frac{\omega_\beta}{\omega_1} \Delta\nu  \\
a_\nu & = & \frac{K_0[\hspace{-0.04cm}J\hspace{-0.09cm}J\hspace{-0.03cm}]\omega_1}{4\gamma_0^2\omega_\beta\rho}e^{-i\Delta\nu\frac{\omega_\beta}{\omega_1}\tau} \nu A_\nu \\
f & = & \rho F
\end{eqnarray}
We thus obtain:
\begin{equation}
\frac{\partial (\nu A_\nu)}{\partial\tau}   =  \frac{(4+K_0^2)\omega_\beta\rho^2}{2K_0[\hspace{-0.04cm}J\hspace{-0.09cm}J\hspace{-0.03cm}]\omega_1} e^{i\Delta\hat{\nu}\hat{\tau}} \left[ \frac{\partial}{\partial\hat{\tau}} + i\Delta\hat{\nu} \right] a_\nu
\end{equation}
So:
\begin{eqnarray}
\left[ \frac{\partial}{\partial\hat{\tau}} + i\Delta\hat{\nu} \right] a_\nu & = & - \frac{I}{I_A} \frac{4K_0^2[\hspace{-0.04cm}J\hspace{-0.09cm}J\hspace{-0.03cm}]^2\omega_1}{(4+K_0^2)\gamma_0r_0^2k_1^2\omega_\beta\rho^2} \times \ \ \nonumber \\
&& \int  \frac{e^{-i\nu\phi}}{2\pi} f \ \! d\phi d\hat{\eta} dK \frac{d\theta_r}{2\pi}  
\end{eqnarray}
By defining $\rho$ (the equivalent of the Pierce parameter in FEL theory) as:
\begin{eqnarray}
\rho & = & \frac{1}{\gamma_0} \left[ \frac{I}{I_A} \frac{2K_0^2(2+K_0^2)^2[\hspace{-0.04cm}J\hspace{-0.09cm}J\hspace{-0.03cm}]^2}{(4+K_0^2)^2r_0^2k_\beta^2} \right]^{1/3} \nonumber \\
     & = & \left[ \frac{I}{I_A} \frac{2(2+K_0^2)^2[\hspace{-0.04cm}J\hspace{-0.09cm}J\hspace{-0.03cm}]^2}{(4+K_0^2)^2\gamma_0} \right]^{1/3}
\label{def_rho}
\end{eqnarray}
we find:
\begin{equation} 
\left[ \frac{\partial}{\partial\hat{\tau}} + i\Delta\hat{\nu} \right] a_\nu  = - \int  \frac{\Delta\nu}{\Delta\hat{\nu}} \frac{e^{-i\nu\phi}}{2\pi} f \ \! d\phi d\hat{\eta} dK \frac{d\theta_r}{2\pi}  
\label{eq_wave_normalized} 
\end{equation}
Moreover, the Vlasov equation becomes:
\begin{eqnarray}
\frac{\partial f}{\partial \hat{\tau}} + \hat{\eta}\frac{\partial f}{\partial \phi} + \left( \int a_\nu e^{i\nu\phi}d\Delta\hat{\nu} + c.c. \right)\frac{\partial f}{\partial\hat{\eta}} && \nonumber \\
+ K'\frac{\partial f}{\partial K} + \theta_r'\frac{\partial f}{\partial \theta_r} & = & 0 \hspace{1.0cm}
\label{normalized_vlasov_eq}
\end{eqnarray}
where $X'=dX/d\hat{\tau}$.

\subsection{Calculation of the growth rate}

To calculate the growth rate, we need to solve the coupled equations (\ref{eq_wave_normalized}) and (\ref{normalized_vlasov_eq}). 

Equation (\ref{normalized_vlasov_eq}) can be linearized in the small signal regime before saturation when the scaled radiation field is small, i.e.:
\begin{equation} 
\int a_\nu e^{i\nu\phi}d\Delta\hat{\nu}+c.c.  =  \hat{\eta}'  \ll 1
\end{equation}
Let us split $f$ in two parts:  
\begin{equation}
f = f_0 + f_1 
\end{equation}
where $f_0$ is the distribution function averaged over $\phi$ and $f_1$ contains the noise fluctuation and the modulation induced by the bunching.
The average over $\phi$ of Eq.(\ref{normalized_vlasov_eq}) leads to:
\begin{eqnarray}
\frac{\partial f_0}{\partial \hat{\tau}} + \left<\left( \int a_\nu e^{i\nu\phi}d\Delta\hat{\nu} + c.c. \right)\frac{\partial f_1}{\partial\hat{\eta}}\right>_\phi && \nonumber \\
+ K'\frac{\partial f_0}{\partial K} + \theta_r'\frac{\partial f_0}{\partial \theta_r} & = & 0  \hspace{0.6cm}
\label{normalized_vlasov_eq_average_phi}
\end{eqnarray}
The small signal regime also implies that $f_1 \ll f_0$. 
We can then assume that the second term in LHS of Eq.~(\ref{normalized_vlasov_eq_average_phi}) can be neglected, which leads to:
\begin{equation}
\frac{\partial f_0}{\partial \hat{\tau}}+ K'\frac{\partial f_0}{\partial K} + \theta_r'\frac{\partial f_0}{\partial \theta_r} = 0 
\label{normalized_vlasov_eq_average_phi_neglect}
\end{equation}
The corresponding equation for $f_1$ is therefore:
\begin{eqnarray}
\frac{\partial f_1}{\partial \hat{\tau}} + \hat{\eta}\frac{\partial f_1}{\partial \phi} + \left( \int a_\nu e^{i\nu\phi}d\Delta\hat{\nu} + c.c. \right)\frac{\partial f_0}{\partial\hat{\eta}} && \nonumber \\
+ K'\frac{\partial f_1}{\partial K} + \theta_r'\frac{\partial f_1}{\partial \theta_r} & = & 0  \hspace{1.0cm}
\label{normalized_vlasov_eq_f1}
\end{eqnarray}

To solve this equation, we consider the trajectory of an electron, which is parameterized by $\phi^{(0)}$, $\hat{\eta}^{(0)}$, $K^{(0)}$ and $\theta_r^{(0)}$.
According to the Vlasov equation (\ref{normalized_vlasov_eq}), we have:
\begin{equation}
\frac{d}{d\hat{\tau}} f(\phi^{(0)},\hat{\eta}^{(0)},K^{(0)},\theta_r^{(0)},s) = 0
\end{equation}
where  $\phi^{(0)}$, $\hat{\eta}^{(0)}$, $K^{(0)}$ and $\theta_r^{(0)}$ are here given at the time $s$.
Thanks to Eq.(\ref{normalized_vlasov_eq_f1}), we can write:
\begin{eqnarray}
&& \frac{d}{d\hat{\tau}} f_1(\phi^{(0)},\hat{\eta}^{(0)},K^{(0)},\theta_r^{(0)},s) =   \nonumber \\
&& \hspace{0.7cm} - \left( \int a_\nu(s) e^{i\nu\phi^{(0)}}d\Delta\hat{\nu} + c.c. \right) \times  \nonumber \\
&& \hspace{1.1cm} \frac{\partial f_0}{\partial\hat{\eta}}(\hat{\eta}^{(0)},K^{(0)},\theta_r^{(0)},s)
\end{eqnarray}
So:
\begin{eqnarray}
&&f_1(\phi,\hat{\eta},K,\theta_r,\hat{\tau})   =   f_1(\phi^{(0)},\hat{\eta}^{(0)},K^{(0)},\theta_r^{(0)},0) \nonumber \\
&& \hspace{0.7cm} - \int_{0}^{\hat{\tau}}  \left( \int a_\nu(s) e^{i\nu\phi^{(0)}}d\Delta\hat{\nu} + c.c. \right)  \nonumber \\
&& \hspace{1.7cm} \times \frac{\partial f_0}{\partial\hat{\eta}}(\hat{\eta}^{(0)},K^{(0)},\theta_r^{(0)},s)  ds
\end{eqnarray}
where $\phi$, $\hat{\eta}$, K and $\theta_r$ are the values of $\phi^{(0)}$, $\hat{\eta}^{(0)}$, $K^{(0)}$ and $\theta_r^{(0)}$ at time $\hat{\tau}$.
Moreover, we have:
\begin{equation}
\phi^{(0)}(s)  =  \phi + \int_{\hat{\tau}}^{s} \hat{\eta}^{(0)}(\check{s}) d\check{s}
\end{equation}
So we find that:
\begin{widetext}
\begin{eqnarray}
\int \frac{\Delta\nu }{\Delta\hat{\nu}} \frac{e^{-i\nu\phi}}{2\pi} f_1(\phi,\hat{\eta},K,\theta_r,\hat{\tau}) d\phi &  = & \int \frac{\Delta\nu }{\Delta\hat{\nu}} \frac{e^{-i\nu\phi}}{2\pi} f_1(\phi^{(0)},\hat{\eta}^{(0)},K^{(0)},\theta_r^{(0)},0) d\phi   \nonumber \\
&& - \int_{0}^{\hat{\tau}} a_\nu(s) \exp\!\left[i\nu\int_{\hat{\tau}}^{s} \hat{\eta}^{(0)}(\check{s}) d\check{s}\right] \frac{\partial f_0}{\partial\hat{\eta}}(\hat{\eta}^{(0)},K^{(0)},\theta_r^{(0)},s) ds  \hspace{0.7cm} \label{eq_int_phi_f1}
\end{eqnarray}
\end{widetext}
In the small signal regime, provided that $\hat{\eta}'\ll1$, we can assume that $ \hat{\eta}^{(0)}(s)\sim\hat{\eta}$. 
From Eq.(\ref{eq_def_K_eta}), we can deduce that $K'=(1+K_0^2)\rho\hat{\eta}'/(2K_0^2)$ so $K'\ll1$ and $K^{(0)}\sim K$.
Based on Eq.(\ref{normalized_vlasov_eq_average_phi_neglect}) and on the definition of $\theta_r^{(0)}$, we can deduce that $f_0(\theta_r^{(0)},s)$ is a constant if we assume that $K'=0$.
Therefore, $f_0(\theta_r^{(0)},s)=f_0(\theta_r,\hat{\tau})$, which leads to:
\begin{widetext}
\begin{equation}
\int_{0}^{\hat{\tau}} a_\nu(s) \exp\!\left[i\nu\int_{\hat{\tau}}^{s} \hat{\eta}^{(0)}(\check{s}) d\check{s}\right] \frac{\partial f_0}{\partial\hat{\eta}}(\hat{\eta}^{(0)},K^{(0)},\theta_r^{(0)},s) ds = \frac{\partial f_0}{\partial\hat{\eta}}(\hat{\eta},K,\theta_r,\hat{\tau})  \int_{0}^{\hat{\tau}} a_\nu(s) e^{i\nu\hat{\eta}(s-\hat{\tau})}  ds  \label{eq_simplification_int_anu_df0_dt}
\end{equation}
\end{widetext}

As $f_0$ does not depend on $\phi$, we have the following result if we assume that the electron beam is very long in comparison to the fundamental radiation wavelength $\lambda_1$:
\begin{equation}
\int e^{-i\nu\phi} f_0 d\phi  \ll  \int e^{-i\nu\phi} f_1 d\phi
\end{equation}
Then, Eq.(\ref{eq_wave_normalized}) becomes:
\begin{equation}
\left[ \frac{\partial}{\partial\hat{\tau}} + i\Delta\hat{\nu} \right] a_\nu  = - \int  \frac{\Delta\nu}{\Delta\hat{\nu}} \frac{e^{-i\nu\phi}}{2\pi} f_1 \ \! d\phi d\hat{\eta} dK \frac{d\theta_r}{2\pi}  
\label{eq_wave_normalized_f1}
\end{equation}
By using Eq.(\ref{eq_int_phi_f1}), (\ref{eq_simplification_int_anu_df0_dt}) and (\ref{eq_wave_normalized_f1}), we obtain:
\begin{widetext}
\begin{equation}
\left[ \frac{\partial}{\partial\hat{\tau}} + i\Delta\hat{\nu} \right] a_\nu   -  \int \frac{\partial f_0}{\partial\hat{\eta}} \int_0^{\hat{\tau}} a_\nu(s) e^{i\hat{\eta}(s-\hat{\tau})} ds  d\hat{\eta} dK \frac{d\theta_r}{2\pi} = - \int  \frac{\Delta\nu}{\Delta\hat{\nu}} \frac{e^{-i\nu\phi}}{2\pi} f_1(\phi^{(0)},\hat{\eta},K,\theta_r^{(0)},0) \ \! d\phi d\hat{\eta} dK \frac{d\theta_r}{2\pi}  
\label{eq_wave_f1_integral}
\end{equation}
\end{widetext}
This equation shows that each frequency component of the radiation field is independently amplified.
The RHS of Eq.(\ref{eq_wave_f1_integral}) corresponds to the initial fluctuation and is the source term that creates the initial radiation in the absence of seed.

To determine the growth rate, we only consider the homogeneous part of Eq.(\ref{eq_wave_f1_integral}).   
We seek a solution in which $a_\nu$ is proportional to $\exp(-i\hat{\mu}\hat{\tau})$, where $\hat{\mu}$ is the complex growth rate.
Then, we have $a_\nu(s)=a_\nu(\hat{\tau})\exp[-i\hat{\mu}(s-\hat{\tau})]$.
This leads to:
\begin{equation}
-i\hat{\mu} + i\Delta\hat{\nu}    -  \int  \frac{\partial f_0}{\partial\hat{\eta}}  \int_0^{\hat{\tau}} e^{i(\hat{\eta}-\hat{\mu})(s-\hat{\tau})} ds  d\hat{\eta} dK \frac{d\theta_r}{2\pi}   = 0
\end{equation}
We first calculate the integral over the time $s$.
Then, we assume that $\eta$, $K$ and $\theta_r$ are not correlated at the time $\tau$.
Thanks to Eq.(\ref{def_F_cas_ideal_part_1}--\ref{def_F_cas_ideal_part_2_D4}), the integration over $K$ and $\theta_r$ leads to:
\begin{equation}
-i\hat{\mu} + i\Delta\hat{\nu}    +i  \int  \frac{\partial f_0(\hat{\eta})}{\partial\hat{\eta}}  \frac{1}{\hat{\eta}-\hat{\mu}}  d\hat{\eta}  = 0
\end{equation}
Here we have also assumed that $|\exp[i\hat{\mu}\hat{\tau}]|\ll1$, as $\exp[-i\hat{\mu}\hat{\tau}]$ is supposed to growth exponentially with time.
After integrating by part over $\hat{\eta}$, we obtain:
\begin{equation}
-i\hat{\mu} + i\Delta\hat{\nu}    +i  \int   \frac{f_0(\hat{\eta})}{(\hat{\eta}-\hat{\mu})^2}  d\hat{\eta}  = 0
\end{equation}
In the limit where there is no energy spread ($f_0(\hat{\eta})=\delta(\hat{\eta})$), this equation becomes:
\begin{equation}
\hat{\mu}^2(\hat{\mu}-\Delta\hat{\nu}) = 1
\label{eq_mu_cube_delta_nu}
\end{equation}
At the optimal frequency ($\Delta\hat{\nu}=0$), we obtain:
\begin{equation}
\hat{\mu}^3  = 1
\end{equation}
The solution with the largest imaginary part is associated to the largest growth rate. 
Thus, we only consider the following solution:
\begin{equation}
\hat{\mu}  = -\frac{1}{2} + i\frac{\sqrt{3}}{2}
\end{equation}
Thanks to Eq.(\ref{def_hat_tau}), we finally find that the field amplitude is proportional to:
\begin{equation}
|a_{\nu=1}(\tau)|  \propto  \exp\left[\frac{\sqrt{3}(4+K_0^2)\rho}{16\gamma_0^2}\tau\right] 
\end{equation}

In the following, the parameter $\rho$ given in the 1D approximation by Eq.(\ref{def_rho}) will be referred as $\rho_{\rm  1D}$.  
The 1D gain time for the field amplitude is then:
\begin{equation}
\tau_{G}^{\rm 1D}  =  \frac{16\gamma_0^2}{(4+K_0^2)\sqrt{3}\rho_{\rm  1D}}\omega_1^{-1}  =  \frac{2(2+K_0^2)}{\pi(4+K_0^2)\sqrt{3}\rho_{\rm  1D}}\tau_\beta
\end{equation}
The associated power or intensity gain length is then:
\begin{equation}
L_{GP}^{\rm 1D}  =  \frac{(2+K_0^2)}{\pi(4+K_0^2)\sqrt{3}\rho_{\rm 1D}}\lambda_\beta
\end{equation}

\section{Transverse effect: influence of the Rayleigh length}
\label{appC}

In Appendix \ref{appB}, we have assumed that the transverse variation of $A_\nu$ can be neglected, as we have used $\boldsymbol{\nabla}_\bot^2=0$ to simplify Eq.(\ref{eq_wave_slow_vary_env}).
However, if we consider that the electron beam creates a radiation beam with a waist close to the electron beam radius $r_0$, then the associated Rayleigh length is $Z_r\sim r_0^2k_1/2$.
This length is much shorter than the gain length, since: 
\begin{equation}
\frac{Z_r}{c\tau_{GI}}  \sim  \frac{(4+K_0^2)K_0^2\sqrt{3}}{(2+K_0^2)^2} \rho
\end{equation}
so $Z_r/(c\tau_{GI}) \ll 1$ as $\rho\ll1$.
Therefore, the intensity of the emitted radiation is strongly reduced after one gain length, which reduces the growth rate.

To take into account this phenomenon, we assume that the current $j_\nu$ generates a Gaussian beam with a waist $W_0=w_0r_0$, where $w_0$ is a free parameter. 
In the following, for the sake of simplicity, we consider only the 2D case.
We have then$\boldsymbol{\nabla}_\bot^2=\partial^2/\partial (k_1x)^2$.
We introduce $A_{\nu,\check{\tau}}(\tau,x)$, which is the field generated by the current $j_\nu(\tau,x)\delta(\tau-\check{\tau})$.
Therefore, $A_{\nu,\check{\tau}}(\tau,x)$ is a solution of:
\begin{equation}
\left(  2i\nu \frac{\partial}{\partial\tau} + \frac{\partial^2}{\partial (k_1x)^2} \right) A_{\nu,\check{\tau}}(\tau,x) = - \frac{4\pi e\omega_1}{I_A} j_\nu(\tau,x) \delta(\tau-\check{\tau})
\label{eq_wave_A_nu_heaviside}
\end{equation}
By assuming that the current $j_\nu$ generates a Gaussian beam with a waist $W_0$, we find that:
\begin{eqnarray}
A_{\nu,\check{\tau}}(\tau,x) & = & A_{\nu,\check{\tau}}(\check{\tau},0)H(\tau-\check{\tau}) B(\tau-\check{\tau},x)  \label{def_A_nu_check_tau} \\
B(\tau,x) & = & \frac{1}{\left( 1+ \frac{\tau^2}{Z_r^2k_1^2} \right)^{1/4}} \times \nonumber \\
&& e^{ \frac{-x^2}{W^2(\tau)}} e^{i\frac{\nu k_1x^2}{2R(\tau)}-\frac{i}{2}\arctan\left(\frac{\tau}{Z_rk_1}\right)}
\end{eqnarray}
where $H$ is the Heaviside function, and:
\begin{eqnarray}
Z_r           & = &  \frac{\nu W_0^2 k_1}{2} \\
W(\tau)   & = &  W_0\sqrt{1+\frac{\tau^2}{Z_r^2k_1^2}} \\
R(\tau)     & = &  \tau\left( 1+\frac{Z_r^2k_1^2}{\tau^2} \right) 
\end{eqnarray}
$B$ is a solution of the 2D paraxial wave equation, so:
\begin{equation}
\left(  2i\nu \frac{\partial}{\partial\tau} + \frac{\partial^2}{\partial (k_1x)^2} \right) B = 0
\label{eq_wave_B}
\end{equation}

Assuming that $A_{\nu,\check{\tau}}(\tau,x)$ is a solution of Eq.(\ref{eq_wave_A_nu_heaviside}) implies that $j_\nu(\tau,x) = j_\nu(\tau,0)\exp(-x^2/W_0^2)$.
$W_0$ (and thus $w_0$) should thus be chosen so that the function $j_\nu(\tau,0)\exp(-x^2/W_0^2)$ provides the best fit of the real $j_\nu$ given by Eq.(\ref{j_nu_integral_form}). 
As a matter of fact, $W_0$ is close to $r_0$ so $w_0$ is close to 1.
Based on those definitions, we can moreover find that:
\begin{equation}
A_{\nu,\check{\tau}}(\check{\tau},0)  =  -\frac{2\pi e\omega_1}{i\nu I_A} j_\nu(\check{\tau},0) 
\label{def_A_nu_check_tau_0}
\end{equation}

By integrating Eq.(\ref{eq_wave_A_nu_heaviside}) over $\check{\tau}$, we obtain:
\begin{eqnarray}
\left(  2i\nu \frac{\partial}{\partial\tau} + \frac{\partial^2}{\partial (k_1x)^2} \right) \int_{-\infty}^{+\infty} A_{\nu,\check{\tau}}(\tau,x) d\check{\tau} = && \nonumber \\
- \frac{4\pi e\omega_1}{I_A} j_\nu(\tau,x) &&
\end{eqnarray}
The identification with Eq.(\ref{eq_wave_slow_vary_env}) shows that:
\begin{equation}
A_\nu(\tau,x)  =  \int_{-\infty}^{+\infty} A_{\nu,\check{\tau}}(\tau,x) d\check{\tau} = \int_{0}^{\tau} A_{\nu,\check{\tau}}(\tau,x) d\check{\tau}
\label{def_A_nu_function_A_nu_tau}
\end{equation}
The limits of the integral can be changed from $(-\infty,+\infty)$ to $(0,\tau)$ because we consider that nothing happens when $\tau<0$ (i.e. $j_\nu=A_{\nu,\check{\tau}}=0$ if $\tau<0$), and thanks to the presence of the function $H$, we have $A_{\nu,\check{\tau}}(\tau,x)=0$ if $\check{\tau}>\tau$. 

As in Appendix \ref{appB}, we seek a solution where the current and field are proportional to $\exp(-i\mu\tau)$, with $\mu$ the complex growth rate. 
The current thus satisfies $j_\nu(\tau,0)=j_\nu(0,0)\exp(-i\mu\tau)$.
According to Eq.(\ref{def_A_nu_check_tau_0}), we get:
\begin{equation}
A_{\nu,\check{\tau}}(\check{\tau},0)  =  A_{\nu,0}(0,0) e^{-i\mu\check{\tau}}
\end{equation}
Based on Eq.(\ref{def_A_nu_check_tau}) and (\ref{def_A_nu_function_A_nu_tau}), we can write:
\begin{eqnarray}
A_\nu(\tau,x)  & = &  \int_{0}^{\tau}  A_{\nu,0}(0,0) e^{-i\mu\check{\tau}} B(\tau-\check{\tau},x) d\check{\tau}  \nonumber \\
               & = & A_{\nu,0}(0,0) e^{-i\mu\tau}  \int_{0}^{\tau} e^{i\mu\check{\tau}} B(\check{\tau},x) d\check{\tau} \hspace{0.4cm} \label{A_nu_function_A_nu_000}
\end{eqnarray}
By using Eq.(\ref{eq_wave_B}), we can then write:
\begin{widetext}
\begin{eqnarray}
\frac{\partial^2 A_\nu(\tau,x)}{\partial(xk_1)^2}  & = &  -2i\nu A_{\nu,0}(0,0) e^{-i\mu\tau}  \int_{0}^{\tau} e^{i\mu\check{\tau}} \frac{\partial B(\check{\tau},x)}{\partial \check{\tau}} d\check{\tau}  \nonumber \\
                                                   & = &  -2\mu\nu A_\nu(\tau,x) +2i\nu A_{\nu,0}(0,0) e^{-i\mu\tau} \left[ B(0,x) - e^{i\mu\tau} B(\tau,x) \right]
\end{eqnarray}
\end{widetext}
where we have also performed an integration by parts over the time variable. 
From Eq.(\ref{A_nu_function_A_nu_000}), we can also  write:
\begin{equation}
\frac{\partial A_\nu(\tau,x)}{\partial\tau} =  -i\mu A_\nu(\tau,x) + A_{\nu,0}(0,0) B(\tau,x) 
\end{equation}
which leads to the following result:
\begin{eqnarray}
\left( 2i\nu\frac{\partial}{\partial\tau} + \frac{\partial^2}{\partial(xk_1)^2}  \right) A_\nu(\tau,x)  =  && \nonumber \\
2i\nu A_{\nu,0}(0,0) e^{-i\mu\tau} B(0,x) &&
\label{eq_wave_temp1}
\end{eqnarray}

To go further, we now assume that the term $\int_{0}^{\tau} e^{i\mu\check{\tau}} B(\check{\tau},x)$ that appears in Eq.(\ref{A_nu_function_A_nu_000}) becomes constant after some time (after few gain times).
Indeed, as we have supposed that $A_\nu$ is exponentially growing, then $e^{i\mu\check{\tau}}$ is exponentially decreasing and the integral stays constant if $\tau\gg1/\Im(\mu)$.
This assumption has been verified numerically: the integral reaches a nearly constant value after few gain times.
We can then write:
\begin{eqnarray}
A_\nu(\tau,x)  =   A_{\nu,0}(0,0) e^{-i\mu\tau}  \int_{0}^{+\infty} e^{i\mu\check{\tau}} B(\check{\tau},x) d\check{\tau}
\end{eqnarray}
So:
\begin{eqnarray}
\frac{\partial A_\nu(\tau,x)}{\partial\tau} & =  & A_{\nu,0}(0,0) e^{-i\mu\tau} \times \nonumber \\
&& \int_{0}^{+\infty} -i\mu e^{i\mu\check{\tau}} B(\check{\tau},x) d\check{\tau}
\end{eqnarray}
If we define the function $\Gamma(x)$, which is constant with time, as follow:
\begin{eqnarray}
\Gamma(x)  =    \int_{0}^{+\infty} -i\mu e^{i\mu\check{\tau}} \frac{B(\check{\tau},x)}{B(0,x)} d\check{\tau}
\end{eqnarray}
then we obtain from Eq.(\ref{eq_wave_temp1}):
\begin{equation}
\left( 2i\nu\frac{\partial}{\partial\tau} + \frac{\partial^2}{\partial(xk_1)^2}  \right) A_\nu(\tau,x)  =  \frac{2i\nu}{\Gamma(x)} \frac{\partial A_\nu(\tau,x)}{\partial\tau}
\end{equation}
Therefore, to calculate the on-axis field amplification we can follow the same method as in Appendix \ref{appB}, but the LHS of Eq.~(\ref{eq:paraxial_eq_1D}) has to be replace by the term $(1/\Gamma)\partial A_\nu/\partial\tau$, where $\Gamma=\Gamma(0)$.
By defining the new Pierce parameter $\rho$ as:
\begin{equation}
\rho  =  \rho_{\rm 1D}|\Gamma|^{\frac{1}{3}}
\label{def_pierce_param_multi_D}
\end{equation}
then Eq.(\ref{eq_wave_normalized}) becomes:
\begin{equation}
\left[ \frac{\partial}{\partial\hat{\tau}} + i\Delta\hat{\nu} \right] a_\nu  = -  e^{i\arg(\Gamma)}  \int  \frac{\Delta\nu}{\Delta\hat{\nu}} \frac{e^{-i\nu\phi}}{2\pi} f \ \! d\phi d\hat{\eta} dK \frac{d\theta_r}{2\pi} 
\end{equation}
We then obtain the equivalent of Eq.(\ref{eq_mu_cube_delta_nu}):
\begin{equation}
\hat{\mu}^2(\hat{\mu}-\Delta\hat{\nu}) =  e^{i\arg(\Gamma)}
\end{equation}
where $\hat{\mu}$ is linked to $\mu$ by:
\begin{equation}
\hat{\mu} =  \frac{8\gamma_0^2}{(4+k_0^2)\rho}\mu
\end{equation}
so that the growth that is given by $e^{-i\mu\tau}$ is also given by $e^{-i\hat{\mu}\hat{\tau}}$.
Eventually, if $\Delta\hat{\nu}=0$, the solution is:
\begin{equation}
\hat{\mu} =  e^{i\left[ \frac{2\pi}{3} + \frac{1}{3}\arg(\Gamma)\right]}
\label{def_mu_hat_multi_D}
\end{equation}
and the gain time associated to the field growth rate is given by:
\begin{equation}
\tau_G =  \frac{8\gamma_0^2}{\Im(\hat{\mu})(4+K_0^2)\rho}
\end{equation}

To be more consistent, we can rewrite $\Gamma$ as a function of $\hat{\mu}$ by changing $\check{\tau}$ into $8\gamma_0^2\check{\tau}/((4+K_0^2)\rho)$:
\begin{eqnarray}
\Gamma =  \int_{0}^{+\infty} -i\hat{\mu} e^{i\hat{\mu}\check{\tau}} B\!\left(\frac{8\gamma_0^2}{(4+K_0^2)\rho}\check{\tau},0\right) d\check{\tau}
\label{def_Gamma_hat_mu}
\end{eqnarray}

The final solution can be found by an iterative method.
We first start from the 1D results $\rho=\rho_{\rm 1D}$ and $\hat{\mu}=e^{2i\pi/3}$ to solve Eq.(\ref{def_Gamma_hat_mu}) and get an approximate value of $\Gamma$.
With this value, we can then solve Eq.(\ref{def_pierce_param_multi_D}) and (\ref{def_mu_hat_multi_D}).
Finally, by solving iteratively those 3 equations, the result found after few loops converges to the solution of those three coupled equations.

Note: this solution is also valid in 3D. 
However, the $B$ function that should be used in 3D is the following:
\begin{eqnarray}
B(\tau,x,y) & = & \frac{1}{\left( 1 + \frac{\tau^2}{Z_r^2k_1^2} \right)^{1/2}} \times \nonumber \\
            &   & e^{ \frac{-(x+y)^2}{W^2(\tau)}} e^{i\frac{\nu k_1(x+y)^2}{2R(\tau)}-i\arctan\left(\frac{\tau}{Z_rk_1}\right)}
\end{eqnarray}
Indeed, the 3D $B$ function should be a solution of the 3D paraxial wave equation:
\begin{equation}
\left(  2i\nu \frac{\partial}{\partial\tau} + \frac{\partial^2}{\partial (k_1x)^2} + \frac{\partial^2}{\partial (k_1y)^2} \right) B = 0 \hspace{0.5cm}
\label{eq_wave_B3D}
\end{equation}

\begin{acknowledgments}
This work was partially supported by FCT (Portugal) through grant PTDC/FIS/111720/2009, and by the European community through LaserLab Europe/CHARPAC EC FP7 Contract No 228464.
The simulations were performed at the IST Cluster (Portugal), and at Jaguar supercomputer under INCITE.
\end{acknowledgments}


\end{document}